\documentclass{amsart}

\usepackage{graphicx}
\usepackage{amsmath}
\theoremstyle{definition}

\theoremstyle{remark}

\numberwithin{equation}{section}

\begin{document}

\pagestyle{empty}

\title{Negative exponential behavior of image mutual information for pseudo-thermal light ghost imaging: Observation, modeling, and verification}

\author{Junhui Li$^1$}
\address{$^1$ State Key Laboratory of Advanced Optical Communication Systems and Networks, School of Electronics Engineering and Computer Science, and Center for Quantum Information Technology, Peking University, Beijing 100871, China}

\author{Bin Luo$^2$}
\address{$^2$ State Key Laboratory of Information Photonics and Optical Communications, Beijing University of Posts and Telecommunications, Beijing 100876, China}

\author{Dongyue Yang$^3$}
\address{$^3$ School of Electronic Engineering, Beijing University of Posts and Telecommunications, Beijing 100876, China}

\author{Longfei Yin$^3$}

\author{Guohua Wu$^3$}

\author{Hong Guo$^1$}
\email{hongguo@pku.edu.cn}



\begin{abstract}
When use the image mutual information to assess the quality of reconstructed image in pseudo-thermal light ghost imaging, a negative exponential behavior with respect to the measurement number is observed. Based on information theory and a few simple and verifiable assumptions, semi-quantitative model of image mutual information under varying measurement numbers is established. It is the Gaussian characteristics of the bucket detector output probability distribution that leads to this negative exponential behavior. Designed experiments verify the model.
\end{abstract}

\maketitle

\section{Introduction}
\label{sec:1}
Image quality assessment is known to be difficult so far \cite{ImageAssessment}. Besides traditional image quality measures based on error estimation, assessments of different types are introduced, first by the signal processing community (cf. \cite{SignalAssessment} for a review). Among others, the mutual information (MI), representing the amount of information shared by two random variables in information theory \cite{InformationTheoryBook}, was introduced to account the similarity between images \cite{IMI1,IMI2}, and has been successfully applied in different circumstances to assess image quality \cite{IMIApp1,IMIApp2}.

Being different from the usual ``single snapshot'' imaging process, ghost imaging (GI) is built on a large number of consecutive measurements on two quantities: light intensity registered by a ``bucket'' detector with no spatial resolution, and a spatial profile that never reaches the object---either an ``idler'' reference light field \cite{Shih95}, a modulation pattern \cite{IEEE08}, or the calculated diffraction profile of that field \cite{Shapiro08}. As a consecutive process, modeling the performance under varying measurement numbers is of great significance to GI. One would naturally expect the image quality to improve with increasing measurement number $n$, and converge when~$n$~is quite large, suggesting an upper limit of image quality when~$n \to \infty$. Unfortunately, previous studies of image quality focus on the influence of either the noise level \cite{SNR02,SNR07,SNR08} or relative spatial/temporal scale \cite{SNR11}, and no quantitative analysis concerning measurement number has been published according to our knowledge, except for a few qualitative observations \cite{IEEE08,IterativeGI} and an untight lower bound \cite{CSLowerBound}.

In this contribution, we use image mutual information (IMI) between the object~$O$~and the reconstructed image~$Y$~to assess image quality of pseudo-thermal light GI. Semi-quantitative fitting shows that IMI~$I\left( {O;Y} \right)$~is a negative exponential function of the measurement number~$n$. An information-theory-based model explains this behavior. All the assumptions are validated. Designed further experiments demonstrate highly agreement with the predictions of the model.

\section{Methods and observation}
\label{sec:2}
\subsection{Experiment setup}
\label{subsec:2.1}
\begin{figure}
\includegraphics[width=\linewidth]{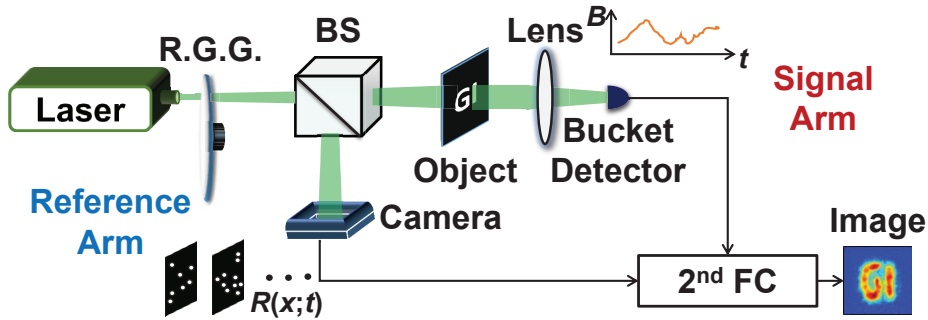}
\caption{Experiment setup. R.G.G. = rotating ground glass. BS = beam splitter. 2$^{\rm{nd}}$ FC = second order fluctuation correlation.}
\label{fig:1}
\end{figure}
A conventional GI setup is implemented as in Fig. \ref{fig:1}. Output of a 532 nm laser passes through a rotating ground glass (R.G.G., Edmund~$100$~mm diameter~$220$~grit gound glass diffuser), turning it into pseudo-thermal light \cite{PseudothermalLight}, whose intensity fluctuates randomly both in the space and time domain. This pseudo-thermal light is then split into two arms by the beam splitter (BS). The signal arm penetrates a transmissive object mask, followed by a focus lens, to be registered as a whole into a temporal intensity sequence~$B\left(t\right)$~by a bucket detector which has no spatial resolution. The spatial profile of the reference arm,~$R\left(x;t\right)$, which never reaches the object, is recorded by a commercial CMOS camera (Thorlabs DCC3240C) synchronically with the bucket detector. The second order fluctuation correlation (2$^{\rm{nd}}$ FC) \cite{PNFC} between corresponding $B\left(t\right)$ and $R\left(x;t\right)$ yields the reconstructed image $Y\left(x\right)$,
\begin{equation}
\label{eq:1}
Y\left( {x} \right) \propto \frac{{{{\left\langle {\left[ {R\left( {x;t} \right) - {{\left\langle {R\left( {x;t} \right)} \right\rangle }_t}} \right] \times \left[ {B\left( t \right) - {{\left\langle {B\left( t \right)} \right\rangle }_t}} \right]} \right\rangle }_t}}}{{{{\left\langle {R\left( {x;t} \right)} \right\rangle }_t}{{\left\langle {B\left( t \right)} \right\rangle }_t}}},
\end{equation}
where $\left\langle  \cdot  \right\rangle_t$ denotes average over all the measurements.

\subsection{Image mutual information}
\label{subsec:2.2}

\textbf{\emph{Mutual information}} In information theory, MI between two random variables~$A$~and~$B$~is defined as
\begin{equation}
\label{eq:2}
I\left( {A;B} \right) = H\left( A \right) - H\left( {A\left| B \right.} \right),
\end{equation}
where $H\left( A \right) =  - \sum\limits_a {{p_A}\left( a \right){{\log }_2}{p_A}\left( a \right)}$ is the Shannon entropy of $A$ with probability distribution function (PDF) ${p_A}\left( a \right)$, denoting the amount of information one reveals when gets full knowledge of ${p_A}\left( a \right)$, and~$H\left( {A\left| B \right.} \right)$ is the conditional entropy of $A$ given $B$, representing the amount of the remain unknown information of $A$ even when the probability distribution of $B$ is totaly determined,
\begin{equation}
\label{eq:3}
H\left( {A\left| B \right.} \right) =  - \sum\limits_a {\sum\limits_b {{p_{A,B}}\left( {a,b} \right){{\log }_2}{p_{A\left| B \right.}}\left( {a\left| b \right.} \right)} } ,
\end{equation}
where ${p_{A,B}}\left( {a,b} \right)$ is the joint probability of $A=a$ and $B=b$, and ${p_{A\left| B \right.}}\left( {a\left| b \right.} \right)$ is the conditional probability of $A=a$ given~$B=b$. Eq. (\ref{eq:2}) shows that $I\left( {A;B} \right)$ denotes the amount of information shared by two partite $A$ and $B$, thus can be a measure of how similar the two variables are, since identical variables have the largest MI, while totally independent ones have the smallest.

\textbf{\emph{Image mutual information}} When IMI is applied, image $A \left( x \right)$ of $N$ pixels is treated as a one-dimensional random variable $A$ of length $N$. IMI between two images $A \left( x \right)$ and $B \left( x \right)$ is defined as MI between two random variables $A$ and $B$, which denotes the similarity between the two images. If $A \left( x \right)$ and $B \left( x \right)$ are set to be the object and image of an imaging system, respectively, IMI can assess the image quality, since the goal of imaging is to accomplish a duplicate as similar to the object as possible. In fact, by maximizing IMI, imaging distortion and relative displacement can be corrected profoundly---known as the image registration technique (cf. \cite{IMIApp1} for a review). Here we want to note that, in order to reduce the influence made by image distortion or relative displacement on the image quality assessment, the image and object should be aligned at first when uses IMI to assess image quality. What is more, unlike other assessments, e.g., mean square error (MSE), IMI is insensitive to the relative coordinate $x$ within the area of interest (AOI), i.e., a rearrangement of pixels, which would totaly change the image content, makes no difference with respect to IMI. This unique property, on one hand, emphasizes the importance of image alignment, and suggests the potential to develop a content-free image quality assessment on the other, which is a general measure of image quality regardless of what pattern it has for a specified image.

\subsection{Observation of negative exponential behavior}
\label{subsec:2.3}
\begin{figure}
\includegraphics[width=\linewidth]{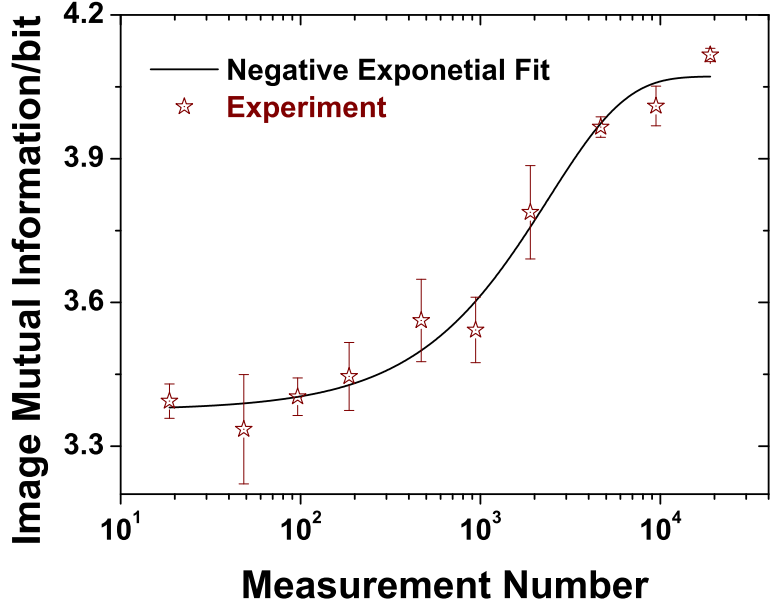}
\caption{Negative exponential behavior. Empty stars are experiment results. Solid line is fitted by Eq. (\ref{eq:4}). Adjusted $R^2=0.97214$ \cite{R2} suggests well fitting.}
\label{fig:2}
\end{figure}

Reconstructed image $Y \left( x \right)$ is recorded against different measurement numbers. The AOI contains $120 \times 120$~pixels. To ensure alignment, the image that is the most over-sampled $\left( n=50000 \right)$, $Y_\infty \left( x \right)$, serves as an almost-identical approximation of the object $O \left( x \right)$, assuming that after so many measurements, the image has been a stable, nearly perfect duplicate to the object. IMI between $O \left( x \right)$ and $Y \left( x \right)$, $I\left( {O;Y} \right)$, is calculated under varying $n$ to assess the image quality---the higher $I\left( {O;Y} \right)$ is, the better quality image one gets. For our system, the image quantization bit length when calculates IMI is set to be $9$, according to the Appendix (Sect. \ref{subsec:A.1}). The result is shown in Fig. \ref{fig:2}. Curve fitting with both linear and nonlinear regression shows that the negative exponential function fits the experiment result best, i.e.,
\begin{equation}
\label{eq:4}
I(O;Y) = {C_1} - {C_2}\exp \left( { - \frac{n}{{{C_3}}}} \right),
\end{equation}
where fitting parameter $C_1$ denotes the upper limit of $I(O;Y)$ when $n \to \infty$, and parameter $C_3$ represents the converge speed, i.e., $C_3$ measurements are required to reduce the uncertainty between image and object to the ${1 \mathord{\left/
 {\vphantom {1 e}} \right. \kern-\nulldelimiterspace} e}$ of its initial value. The larger $C_3$ is, the more measurements one needs to achieve the same level of image quality.

\section{Modeling}
\label{sec:3}
\subsection{GI equivalent model}
\label{subsec:3.1}
\begin{figure}
\includegraphics[width=\linewidth]{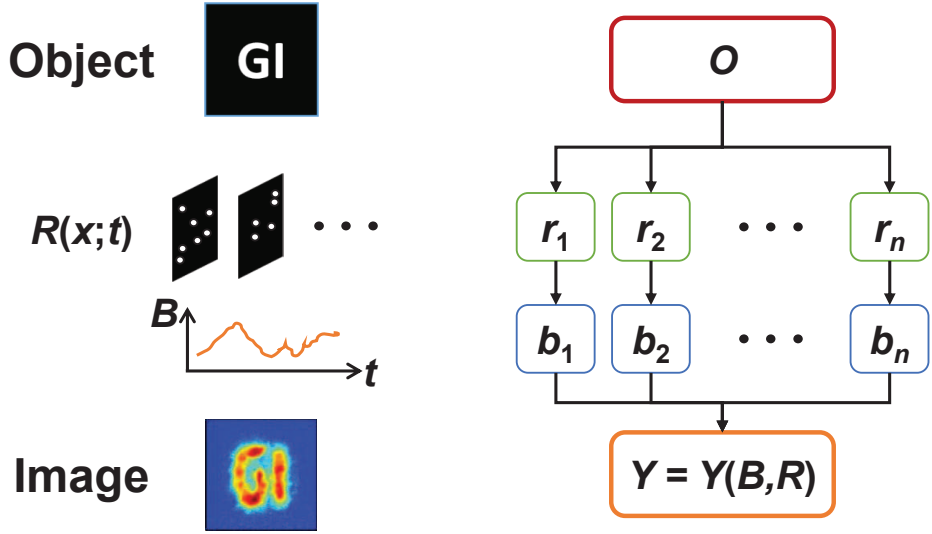}
\caption{GI equivalent model. (Left) GI calculation process. (Right) Corresponding data processing model.}
\label{fig:3}
\end{figure}

In order to explain Eq. (\ref{eq:4}), an equivalent model of GI concerning only the data processing part is established, as shown in Fig. \ref{fig:3}. Light field in the signal arm penetrates the object to be registered by the bucket detector. Under a classical approximation, this light field is identical to its counterpart in the reference arm, up to a propagator, since they are split from the same pseudo-thermal light before the beam splitter in Fig. \ref{fig:1}. Therefore, in each measurement, $O \left( x \right)$ is ``encoded'' by the current reference light field $R \left( {x;t} \right)$ into the bucket detector output $B \left( t \right)$. It is the $n$ realizations of $R \left( {x;t} \right)$ and $B \left( t \right)$, i.e., ${r_1},{r_2}, \ldots {r_n}$ and ${b_1},{b_2}, \ldots {b_n}$, respectively, that lead to the reconstructed image $Y\left( x \right)$ through second order fluctuation correlation of $R \left( {x;t} \right)$ and $B \left( t \right)$, i.e., $Y = Y\left( {\left\{ {{b_n}} \right\},\left\{ {{r_n}} \right\}} \right)$. Therefore, any quantity concerning $O \left( x \right)$ and $Y \left( x \right)$ should be fully determined by random variables $O$, $\left\{ {b_n} \right\}$ and $\left\{ {r_n} \right\}$.

\subsection{Assumptions}
\label{subsec:3.2}
Based on the GI equivalent model, derivation of Eq. (\ref{eq:4}) is shown in the Appendix (Sect.\ref{subsec:A.2}). Here we want to summary and double check all the involving assumptions.

\textbf{\emph{Fixed object}} This assumes that object $O \left( x \right)$ is fixed, so that $H\left( O \right)$ is a constant. This is true since we use a static object, and the most over-sampling image $Y_\infty \left( x \right)$ is a good-enough approximation of $O \left( x \right)$.

\textbf{\emph{Independent object}} This assumes that object $O \left( x \right)$ is independent of the image $Y \left( x \right)$, so that conditional probability ${p_{O\left| Y \right.}}\left( {o\left| y \right.} \right) = {p_O}\left( o \right)$ for all $o$ and $y$. This is true since the object is fixed, no matter what the image is. What is more, since $Y = Y\left( {\left\{ {{b_n}} \right\},\left\{ {{r_n}} \right\}} \right)$, one has ${p_{O\left| Y \right.}}\left( {o\left| Y \right.} \right) = {p_{O\left| {B,R} \right.}}\left( {o\left| {\left\{ {{b_n}} \right\},\left\{ {{r_n}} \right\}} \right.} \right) = {p_O}\left( o \right)$ for any $o$, $Y$, and $\left\{ {{b_n}} \right\}$, $\left\{ {{r_n}} \right\}$, that the fixed object would not be affected by the value of bucket detector nor the reference light field.

\textbf{\emph{Independent reference light field}} This assumes that reference light field $R \left( {x;t} \right)$ is independent of the bucket detector output $B \left( t \right)$, so that conditional probability ${p_{R\left| B \right.}}\left( {r\left| b \right.} \right) = {p_R}\left( r \right)$ for all $b$ and $r$. This is true due to the random nature of the reference light field, both in spatial and time domain, caused by the pseudo-thermal property. We calculate the normalized mutual information (NMI) between the time sequences of bucket detector output $\left\{ {{b_n}} \right\}$ and reference light field $\left\{ {{r_n}} \right\}$ (as for the value of $r_n$, both the summation over the whole AOI and the intensity in a fixed pixel are applied) to verify this independence
\begin{equation}
\label{eq:5}
{I_{nor}}\left( {B;R} \right) = \frac{{2I(B;R)}}{{H\left( B \right) + H\left( R \right) - I(B;R)}},
\end{equation}
where the denominator is the total amount of information, or joint entropy, of $\left\{ {{b_n}} \right\}$ and $\left\{ {{r_n}} \right\}$. Two totally correlated random variables have NMI of unity, while totally independent ones get zero. For varying measurement number $n$, $\left| {{I_{nor}}\left( {B;R} \right)} \right| \le {10^{ - 14}}$, indicating the independence of $\left\{ {{r_n}} \right\}$ on $\left\{ {{b_n}} \right\}$.

\begin{figure}
\includegraphics[width=\linewidth]{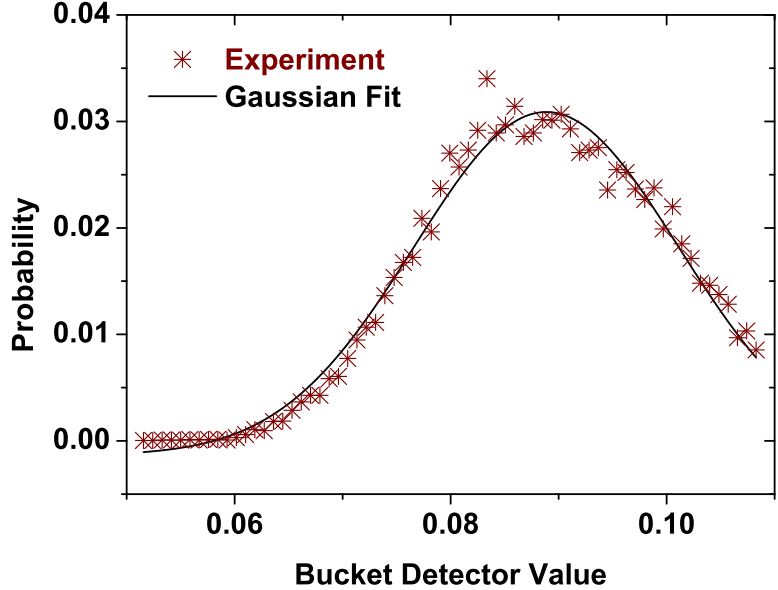}
\caption{Gaussian PDF bucket detector output. Asterisks are experiment results, and solid line is the Gaussian fitting. Adjusted $R^2=0.98041$.}
\label{fig:4}
\end{figure}

\textbf{\emph{Gaussian PDF bucket detector output}} This assumes the output of bucket detector has a Gaussian probability distribution function,
\begin{equation}
\label{eq:6}
{p_B}\left( {{b_i}} \right) \propto \exp \left[ {\frac{{ - {{\left( {{b_i} - \bar b} \right)}^2}}}{{2{\sigma ^2}}}} \right],{\rm{    }}i = 1,2, \ldots ,n,
\end{equation}
where $\bar b$ and $\sigma$ are the expectation and standard derivation of $\left\{ {{b_n}} \right\}$, respectively. The strict derivation for a general case is complex, since specific object pattern is involved. The experiment result is shown in Fig. \ref{fig:4}. Similar result has been reported in Ref. \cite{CorrespondenceGI}.

\textbf{\emph{Bucket detector output iid.}} This assumes that the bucket detector output values are independent and identical distributed (iid.), so that ${p_B}\left( {\left\{ {{b_n}} \right\}} \right) = \mathop {\rm{\Pi }}\limits_{i = 1}^n {p_B}\left( {{b_i}} \right)$. This is guaranteed by the pseudo-thermal nature of the reference light field. Since $R \left( {x;t} \right)$ is random in both the space and time domain \cite{PseudothermalLight}, and ${b_i} = {b_i}\left( {O,{r_i}} \right)$ for any $i \in \left[ {1,n} \right]$, as mentioned in Sect. \ref{subsec:3.1}, the iid. property of $\left\{ {{b_n}} \right\}$ is natural.

\textbf{\emph{Large measurement number}} This assumes $n \gg 1$, which is fulfilled in our experiment.

\textbf{\emph{Image alignment}} This assumes alignment of image to the object. Instead of the real object pattern, we use the most over-sampled image of the same consecutive measurement sequence as the approximation, which ensures perfect alignment as long as the imaging system is stable. Image registration is also conducted. This assumption is not necessary to derive Eq. (\ref{eq:4}), but is vital for IMI to be a good image quality assessment as mentioned in Sect. \ref{subsec:2.2}.

\subsection{Further verification}
\label{subsec:3.3}
Experiments are designed to verify the above model from different perspectives.

\begin{figure*}
\includegraphics[width=\linewidth]{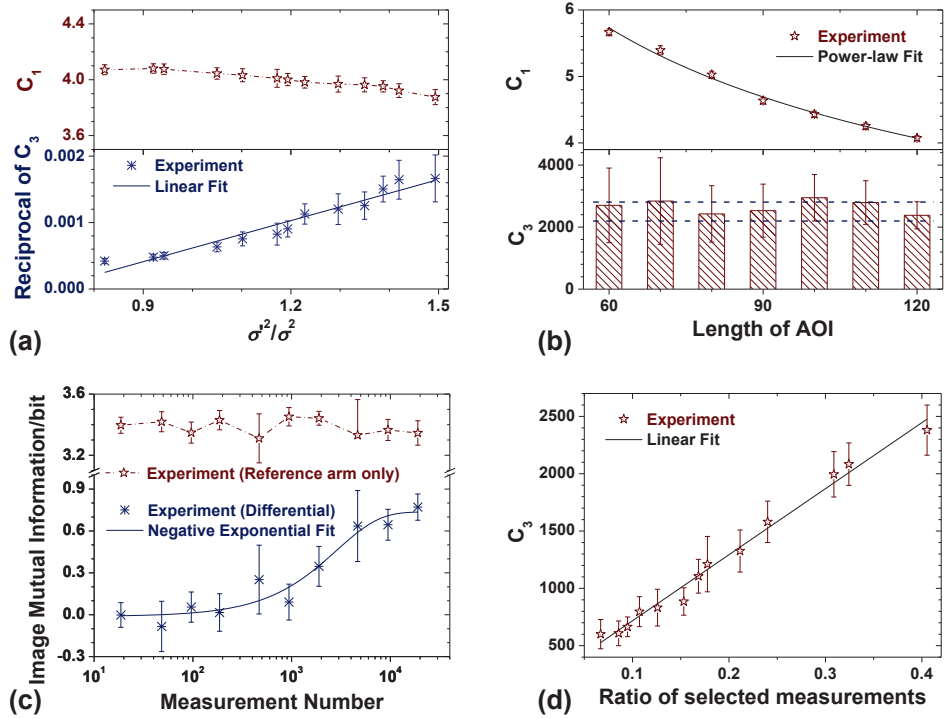}
\caption{Further experimental verifications: (a) Fitting parameters of Eq. (\ref{eq:4}) with bucket fluctuation post-selection: (Up) ${C_1} = {\rm{const}}{\rm{.}}$ in a tolerant error range; (Down) $C_3^{ - 1} \propto {\sigma ^2}$. Asterisks are experiment results, and solid line is the linear fitting by Eq. (\ref{eq:9}). Adjusted $R^2=0.93080$. (b) Fitting parameters of Eq. (\ref{eq:4}) against AOI. (Up) ${C_1} \propto {l^{ - 0.5}}$. Empty stars are experiment results, solid line fitted according to Eq. (\ref{eq:11}). Adjusted $R^2=0.99022$; (Down) ${C_3} = {\rm{const}}{\rm{.}}$ in a tolerant error range, bars are measured results, two dash lines denote the upper and lower bound of $C_3$. (c) Reference-field-only \& Differential IMI. Replacing bucket detector output by reference field intensity summation, in a tolerant error range, reference-field-only IMI has no contribution to the negative exponential behavior, which is dominated by the differential IMI defined in Eq. (\ref{eq:12}). For ${I_{diff}}\left( {O;Y} \right)$ (asterisks) fitted by Eq. (\ref{eq:4}) (solid line, adjusted $R^2=0.92508$). (d) Post-selection on bucket fluctuation cannot reduce total number of conducted measurements. $C_3$ denotes number of measurement required, the ratio represents how many measurements can be selected. Empty stars are experiment results, well fitted by linear function (solid line, adjusted $R^2=0.98611$). The number of measurement required is proportional to the number can be selected with the total number remains unchanged.}
\label{fig:5}
\end{figure*}

\textbf{\emph{Post-selection on bucket fluctuation}} Derivation in the Appendix (Sect. \ref{subsec:A.2}) uses property
\begin{equation}
\label{eq:7}
\mathop {\lim }\limits_{n \to \infty } \frac{{\sum\limits_{i = 1}^n {{{\left( {{b_i} - \bar b} \right)}^2}} }}{n} = {\sigma ^2},
\end{equation}
which is true for all the $\left\{ {{b_n}} \right\}$ as a whole. If one picks up the measurements whose bucket detector output ${b_i}'$ satisfying ${\left( {{b_i}' - \bar b} \right)^2} \ge {m_1}{\sigma ^2}$, i.e., those with larger fluctuations away from the mean value $\bar b$. The variance of selected ${b_i}'$ satisfies
\begin{equation}
\label{eq:8}
\sigma {'^2} = \frac{{\sum\limits_{i = 1}^{n'} {{{\left( {{b_i}' - \bar b} \right)}^2}} }}{{n'}} = {m_2}{\sigma ^2},
\end{equation}
in which $m_2> 1$. Back to Eq. (\ref{eq:4}), one has ${C_3}' = {{{C_3}} \mathord{\left/ {\vphantom {{{C_3}} {{m_2}}}} \right. \kern-\nulldelimiterspace} {{m_2}}}$, which indicates that if the average fluctuation of selected measurements increases, required measurement number for IMI to converge, i.e., to get an image of same quality, decreases with the equal proportion. In other words, one can achieve the same image quality with fewer measurements after post-selection on the fluctuation amplitude of bucket detector output. Similar phenomena has been reported in Ref. \cite{CorrespondenceGI}, and here we present a quantitative relationship this post-selection process should obey
\begin{equation}
\label{eq:9}
\sigma {'^2} \propto \frac{1}{{{C_3}'}}.
\end{equation}
At the same time, according to Sect. \ref{subsec:A.2} and the fixed object approximation, no matter how $m_1$ varies,
\begin{equation}
\label{eq:10}
{C_1} = H\left( O \right) = {\rm{const}}{\rm{.}}
\end{equation}
Experiment results are shown in Fig. \ref{fig:5}(a). The good agreement with Eq. (\ref{eq:9}) and Eq. (\ref{eq:10}) further validates our model.

\textbf{\emph{Varying AOI}} Ref. \cite{SNR11} reported that for a thermal light GI, the ultimate imaging signal-to-noise ratio (SNR) after many measurements should be a rational function of the size of AOI, approximately inversely proportional to the square root of AOI scale parameter $l$. We vary the AOI by choose the $l \times l$ pixels in the middle of the recorded reference light field only in each measurement, and calculate corresponding IMI against measurement number $n$---the same as in Sect. \ref{subsec:2.3}. Experiment shows that Eq. (\ref{eq:4}) holds for all the different AOIs. The results are shown in Fig. \ref{fig:5}(b). Fitting parameter $C_1$, which stands for the ultimate IMI when $n \to \infty$, fits a power-law function of the AOI length $l$ (equivalent to the ``resolution parameter $R$'' in Ref. \cite{SNR11}) well,
\begin{equation}
\label{eq:11}
{C_1} \propto {l^a},
\end{equation}
where $a =  - {\rm{0}}{\rm{.49352}} \pm {\rm{0}}{\rm{.01991}}$. This result agrees with Ref. \cite{SNR11} nicely, suggesting that IMI has very similar behavior with a usual image quality assessment---imaging SNR. What is more, according to our model, the converge parameter $C_3$ should be irrelevant to the size of AOI---which is sort of counter-intuitive---indicating that the number of measurement required before the image quality converges and the image gets stable is the same for both large and small images. However, experiment verifies this conjecture, thus further validates our model.

\textbf{\emph{Contribution of reference field}} Our model suggests that the reference light field has no direct contribution to the negative exponential behavior of the IMI vs. $n$ relationship. To verify this, a comparative experiment is designed that bucket detector output is replaced by summation of reference light field intensities within AOI in each measurement. We calculate IMI between the object and images reconstructed this way, ${Y_{rf}}\left( x \right)$, against measurement number $n$, as shown in Fig. \ref{fig:5}(c). No explicit change of IMI under varying $n$ has been found, which supports our argument. Furthermore, we define ``differential'' IMI as
\begin{equation}
\label{eq:12}
{I_{diff}}\left( {O;Y} \right) = I\left( {O;Y} \right) - I\left( {O;{Y_{rf}}} \right),
\end{equation}
which is the difference between the ordinary IMI and the IMI accomplished by replacing bucket detector output by reference field intensity summation. We expect ${I_{diff}}\left( {O;Y} \right)$ to represent the part of IMI revealed ``solely'' by the bucket detector, and the name ``differential'' comes from the differential entropy in information theory \cite{InformationTheoryBook}. ${I_{diff}}\left( {O;Y} \right)$ also satisfies Eq. (\ref{eq:4}), dominating the negative exponential behavior with one more advantage over $I \left( {O;Y} \right)$---it has a zero initial value when $n \to 0$, making it closer to practical applications since one gets no knowledge of the object before conducting any measurements. This makes ${I_{diff}}\left( {O;Y} \right)$ another promising image quality assessment.

\section{Discussion and conclusion}
\label{sec:4}

\textbf{\emph{Comparison with correspondence GI}} Results of bucket fluctuation post-selection in Sect. \ref{subsec:3.3} are quite similar to those reported in the correspondence GI \cite{CorrespondenceGI}. For example, using only the measurements with large fluctuations can reduce the number of measurement involved in the reconstructed image calculation, and the larger fluctuations are, the fewer measurements are needed. Rather than a coincident, we believe this is due to the same dynamics behind, even though the image calculation formula is different. In both systems, the bucket detector output has a Gaussian PDF, which decreases as the fluctuation increases. In our case, this means measurements with larger fluctuations from the mean bucket detector output value are rarer, and in the language of information theory, one can reveal more information when these measurements appear. We think this explanation also applies to their case. Furthermore, Fig. \ref{fig:5}(d) shows the relationship between converge parameter $C_3$, which denotes the number of measurement required in the bucket fluctuation post-selection calculation, and the ``effective'' ratio---how many measurements in all that being complemented can fulfill the varying fluctuation requirements. Fitting result suggests a linear relationship. Since the total number of measurement stays the same, one can see that reduction of required measurement number is merely due to that fewer measurements can meet the harsher fluctuation requirement---which makes the satisfying measurements rarer, and cannot reduce the total number of measurement to be conducted at the first place. Therefore, correspondence GI can only reduce the number of post-selection measurement, meanwhile still needs the same number of measurement to be complemented.

\textbf{\emph{Applicable range}} One may think that the above negative exponential behavior of image quality versus measurement number relationship is limited to situations with Gaussian PDF or take mutual information as the image quality assessment. Some of our very recent contributions (arXiv: $1603.00371$, $1604.02515$, $1702.08687$) suggest the other way. Our model applies to a quite general scenario thanks to the fundamentality of information theory.

\textbf{\emph{Significance}} Our work is important to the GI community in several ways. First, we provide a semi-quantitative model of the image quality versus measurement number relationship, enabling accurate prediction of the expected image quality after certain number of measurement, and the necessary measurement number to meet any given image quality requirement. Second, explicit connections between GI parameters and concepts in information theory may lead to a new perspective of interdisciplinary research, which we hope could benefit the comparative new and less developed GI study in return. Last but not least, the similar behavior of image mutual information with usual image quality assessments, e.g., mean square error, contrast-to-noise ratio, imaging signal-to-noise ratio, meanwhile the unique insensitivity to specified coordinate within the AOI, suggests that IMI (and differential IMI) is a promising candidate for content-free image quality assessment.

In conclusion, we report the observation, modeling, and verification of the negative exponential behavior when use image mutual information to assess the image quality of pseudo-thermal light ghost imaging, with respect to the measurement number. Rooted in the fundamental information theory, the model applies to a much more general scenario.

\section{Appendix}
\label{sec:Appendix}
\subsection{Number of image quantization levels}
\label{subsec:A.1}
\begin{figure}
\includegraphics[width=\linewidth]{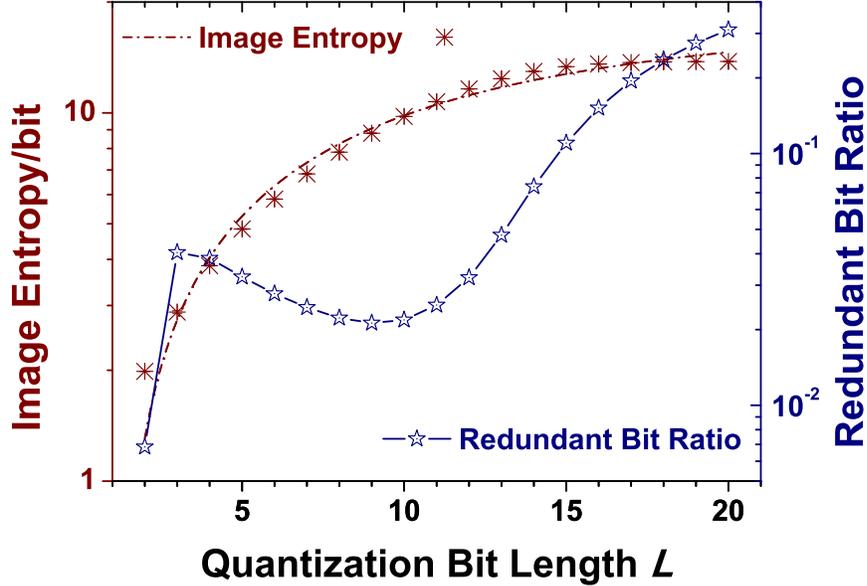}
\caption{Decision of image quantization bit length $L$. Though appears to converge, the image entropy (asterisks \& dashed dot line) grows with ever increasing bit length. Except the leftmost point (binary quantization), the redundant bit ratio (empty stars \& solid line) gets its minimum when $L=9$.}
\label{fig:6}
\end{figure}

When calculates the Shannon entropy or mutual information of (quasi-)continuous random variables, discretization is a practical problem to face with, i.e., to decide how many bits should be used in analog-to-digital conversion. We quantize the the most over-sampled reconstructed image ${Y_\infty} \left( x \right)$, which we used as the approximate object profile in Sect. \ref{subsec:2.3}, into $2^L$ equally-spaced levels, and calculate its image entropy $H\left( {Y_\infty} \right)$ under varying quantization bit length $L$, to decide the optimum $L$, as shown in Fig. \ref{fig:6}.
Unfortunately, the ever growing $H \left( {Y_\infty} \right)$ provides no optimum $L$ except $L \to \infty$, which means that the more bits one uses, the more information one can reveal from the quantized image. We turn to the economy consideration. We define the redundant bit ratio as
\begin{equation}
\label{eq:13}
{R_{bit}} = \frac{{L - H\left( {{Y_\infty }} \right)}}{L},
\end{equation}
which is the ratio of ``redundant'' bits in all $L$ bits when apply entropy coding in the quantization process \cite{InformationTheoryBook}. Small $R_{bit}$ means that it is nearly impossible to compress the quantization bit length, in other words, few are wasted in the $L$ bits. The $R_{bit}$ vs. $L$ relation is also given in Fig. \ref{fig:6}, from which we found that $9$ is the optimum quantization bit length.

\subsection{Derivation of Eq. (\ref{eq:4})}
\label{subsec:A.2}
According to Eq. (\ref{eq:2}),
\begin{equation}
\label{eq:14}
I\left( {O;Y} \right) = H\left( O \right) - H\left( {O\left| Y \right.} \right).
\end{equation}
$H\left( O \right)$ equals to a constant $C_1$ according to the fixed object assumption. Following Eq. (\ref{eq:3}) and the independent object assumption,
\begin{equation}
\label{eq:15}
\begin{array}{rcl}
H\left( {O\left| Y \right.} \right) &=& - \sum\limits_o {\sum\limits_y {{p_{O,Y}}\left( {o,y} \right){{\log }_2}{p_{O\left| Y \right.}}\left( {o\left| y \right.} \right)} }  \\
&=&  - \sum\limits_o {\left[ {{{\log }_2}{p_O}\left( o \right) \cdot \sum\limits_y {{p_{O,Y}}\left( {o,y} \right)} } \right]} ,
\end{array}
\end{equation}
which suggests that the two-fold summation in Eq. (\ref{eq:15}) can be done in two steps: first $\sum\limits_y {p_{O,Y}\left( {o,y} \right)}$ over all the possible values $y$ of the image $Y \left( x \right)$, then over all $o$. Notice that the second step is irrelevant to any specified $y$, and by definition,
\begin{equation}
\label{eq:16}
\sum\limits_y {p_{O,Y}\left( {o,y} \right)}  = p_{O,Y}(o,Y).
\end{equation}
The equivalent GI model in Sect. \ref{subsec:3.1} suggests $Y = Y\left( {\left\{ {{b_n}} \right\},\left\{ {{r_n}} \right\}} \right)$. Together with the independent object assumption, one has
\begin{equation}
\label{eq:17}
\begin{array}{rcl}
{p_{O,Y}}\left( {o,Y} \right) &=& {p_{O,B,R}}\left( {o,\left\{ {{b_n}} \right\},\left\{ {{r_n}} \right\}} \right)\\
 &=& {p_{O\left| {B,R} \right.}}\left( {o\left| {\left\{ {{b_n}} \right\},\left\{ {{r_n}} \right\}} \right.} \right){p_{B,R}}\left( {\left\{ {{b_n}} \right\},\left\{ {{r_n}} \right\}} \right)\\
 &=& {p_O}\left( o \right){p_{B,R}}\left( {\left\{ {{b_n}} \right\},\left\{ {{r_n}} \right\}} \right).
\end{array}
\end{equation}
The independent reference light field assumption leads to
\begin{equation}
\label{eq:18}
\begin{array}{rcl}
{p_{B,R}}\left( {\left\{ {{b_n}} \right\},\left\{ {{r_n}} \right\}} \right) &=& {p_B}\left( {\left\{ {{b_n}} \right\}} \right){p_{R\left| B \right.}}\left( {\left\{ {{r_n}} \right\}\left| {\left\{ {{b_n}} \right\}} \right.} \right)\\
 &=& {p_B}\left( {\left\{ {{b_n}} \right\}} \right){p_R}\left( {\left\{ {{r_n}} \right\}} \right),
\end{array}
\end{equation}
in which ${p_R}\left( {\left\{ {{r_n}} \right\}} \right)$ should be a constant due to the spatial and temporal randomness of the reference light field because of its pseudo-thermal characteristics \cite{PseudothermalLight}. Under the Gaussian PDF and iid. assumptions of the bucket detector output,
\begin{equation}
\label{eq:19}
\begin{array}{rcl}
{p_B}\left( {\left\{ {{b_n}} \right\}} \right) &=& \mathop {\rm{\Pi }}\limits_{i = 1}^n {p_B}\left( {{b_i}} \right)\\
 &\propto& \mathop {\rm{\Pi }}\limits_{i = 1}^n \left\{ {\exp \left[ { - \frac{{{{\left( {{b_i} - \bar b} \right)}^2}}}{{2{\sigma ^2}}}} \right]} \right\}\\
 &=& \exp \left[ { - \frac{{\sum\limits_{i = 1}^n {{{\left( {{b_i} - \bar b} \right)}^2}} }}{{2{\sigma ^2}}}} \right].
\end{array}
\end{equation}
Noticing Eq. (\ref{eq:7}), when the measurement number is large, i.e., $n \gg 1$,
\begin{equation}
\label{eq:20}
\sum\limits_{i = 1}^n {{{\left( {{b_i} - \bar b} \right)}^2}}  = n{\sigma ^2}.
\end{equation}
Substituting Eq. (\ref{eq:16}) to (\ref{eq:20}) into Eq. (\ref{eq:15}),
\begin{equation}
\label{eq:21}
\begin{array}{rcl}
H\left( {O\left| Y \right.} \right) &=&  - \sum\limits_o {\left[ {{p_O}\left( o \right){{\log }_2}{p_O}\left( o \right) \cdot {C_4}\exp \left( { - \frac{n}{{{C_3}}}} \right)} \right]} \\
 &=& H\left( O \right) \cdot {C_4}\exp \left( { - \frac{n}{{{C_3}}}} \right).
\end{array}
\end{equation}
Back to Eq. (\ref{eq:14}), one gets Eq. (\ref{eq:4}). The above derivation shows that it is the probability distribution of bucket detector output that determines the negative exponential behavior of IMI regarding measurement number $n$, which is reasonable, given a fixed object that never changes, a random reference light field which is basically irrelevant to the object, and an iid. bucket output with Gaussian PDF that contains all the information passed to the image from the object.

\subsection*{Conflict of interest}
The authors declare that they have no conflict of interest.

\subsection*{Funding}
National Natural Science Foundation of China (61631014, 61401036, 61471051, 61531003); National Science Fund for Distinguished Young Scholars of China (61225003); China Postdoctoral Science Foundation (2015M580008); Youth Research and Innovation Program of BUPT (2015RC12).

\bibliographystyle{amsplain}

\end{document}